\documentclass[12pt,preprint]{emulateapj}
  
\usepackage{apjfonts}
\shorttitle{Ly$\alpha$ Sizes at z $\sim$ 4.4}
\shortauthors{Finkelstein et al.}

\newcommand{\lya}{Ly$\alpha$}

\newcommand{\fig}[1]{Figure~\ref{#1}}
\def\arcs{\hbox{$^{\prime\prime}$}}

\begin{document}
\slugcomment{Submitted to the Astrophysical Journal}
\title{{\it Hubble Space Telescope} Imaging of Lyman Alpha Emission at \MakeLowercase{z} $\approx$ 4.4\altaffilmark{1}}

\author{Steven   L.    Finkelstein\altaffilmark{2}, Seth H. Cohen\altaffilmark{3}, Rogier A. Windhorst\altaffilmark{3}, Russell E. Ryan\altaffilmark{4}, Nimish P. Hathi\altaffilmark{5}, Keely   D.    Finkelstein\altaffilmark{2}, Jay Anderson\altaffilmark{6}, Norman A. Grogin\altaffilmark{6}, Anton M. Koekemoer\altaffilmark{6}, Sangeeta Malhotra\altaffilmark{3}, Max Mutchler\altaffilmark{6}, James E. Rhoads\altaffilmark{3}, Patrick~J.~McCarthy\altaffilmark{7}, Robert~W.~O'Connell\altaffilmark{8}, Bruce~Balick\altaffilmark{9}, Howard~E.~Bond\altaffilmark{6}, Daniela~Calzetti\altaffilmark{10}, Michael~J.~Disney\altaffilmark{11}, Michael~A.~Dopita\altaffilmark{12},  Jay~A.~Frogel\altaffilmark{13}, Donald~N.~B.~Hall\altaffilmark{14}, Jon~A.~Holtzman\altaffilmark{15}, Randy~A.~Kimble\altaffilmark{16}, Gerard~Luppino\altaffilmark{14},  Francesco~Paresce\altaffilmark{17}, Abhijit~Saha\altaffilmark{18},  Joseph~I.~Silk\altaffilmark{19}, John~T.~Trauger\altaffilmark{20},  Alistair~R.~Walker\altaffilmark{21},  Bradley~C.~Whitmore\altaffilmark{6} \& Erick~T.~Young \altaffilmark{22}}   
\altaffiltext{1}{Based on observations made with the NASA/ESA Hubble Space Telescope, obtained at the Space Telescope Science Institute, which is operated by the Association of Universities for Research in Astronomy, Inc., under NASA contract NAS 5-26555. These observations are associated with program \#11359.}
\altaffiltext{2}{George P. and Cynthia Woods Mitchell Institute for Fundamental Physics and Astronomy, Department of Physics \& Astronomy, Texas A\&M University, College Station, TX 77843; stevenf@physics.tamu.edu}
\altaffiltext{3}{School of Earth and Space Exploration,  Arizona  State University,  Tempe, AZ  85287-1404} 
\altaffiltext{4}{Department of Physics, University of California,  Davis, CA 92616}
\altaffiltext{5}{Department of Physics and Astronomy, University of California,  Riverside, CA 92521}
\altaffiltext{6}{Space Telescope Science Institute, 3700 San Martin Drive, Baltimore, MD 21218}
\altaffiltext{7}{Observatories of the Carnegie Institution of Washington, Pasadena, CA 91101, USA}
\altaffiltext{8}{Department of Astronomy, University of Virginia, Charlottesville, VA 22904-4325}
\altaffiltext{9}{Department of Astronomy, University of Washington, Seattle, WA 98195-1580}
\altaffiltext{10}{Department of Astronomy, University of Massachusetts, Amherst, MA 01003}
\altaffiltext{11}{School of Physics and Astronomy, Cardiff University, Cardiff CF24 3AA, United Kingdom}
\altaffiltext{12}{Research School of Astronomy \& Astrophysics, The Australian National University, ACT 2611, Australia}
\altaffiltext{13}{Association of Universities for Research in Astronomy, Washington, DC 20005}
\altaffiltext{14}{Institute for Astronomy, University of Hawaii, Honolulu, HI 96822}
\altaffiltext{15}{Department of Astronomy, New Mexico State University, Las Cruces, NM 88003}
\altaffiltext{16}{NASA-Goddard Space Flight Center, Greenbelt, MD 20771}
\altaffiltext{17}{Istituto di Astrofisica Spaziale e Fisica Cosmica, INAF, Via Gobetti 101, 40129, Bologna, Italy}
\altaffiltext{18}{National Optical Astronomy Observatories, Tucson, AZ 85726-6732}
\altaffiltext{19}{Department of Physics, University of Oxford, Oxford OX1 3PU, United Kingdom}
\altaffiltext{20}{NASA-Jet Propulsion Laboratory, Pasadena, CA 91109}
\altaffiltext{21}{Cerro Tololo Inter-American Observatory, La Serena, Chile}
\altaffiltext{22}{NASA-Ames Research Center, Moffett Field, CA 94035}

\begin{abstract}
We present the highest redshift detections of resolved \lya\ emission, using {\it Hubble Space Telescope}/ACS F658N narrowband-imaging data taken in parallel with the Wide Field Camera 3 Early Release Science program in the GOODS CDF-S.  We detect \lya\ emission from three spectroscopically confirmed $z =$ 4.4 Ly$\alpha$ emitting galaxies (LAEs), more than doubling the sample of LAEs with resolved \lya\ emission.  Comparing the light distribution between the rest-frame ultraviolet continuum and narrowband images, we investigate the escape of \lya\ photons at high redshift.  While our data do not support a positional offset between the \lya\ and rest-frame ultraviolet (UV) continuum emission, the half-light radii in two out of the three galaxies are significantly larger in \lya\ than in the rest-frame UV continuum.  This result is confirmed when comparing object sizes in a stack of all objects in both bands.  Additionally, the narrowband flux detected with {\it HST} is significantly less than observed in similar filters from the ground.  These results together imply that the \lya\ emission is not strictly confined to its indigenous star-forming regions.  Rather, the \lya\ emission is more extended, with the missing {\it HST} flux likely existing in a diffuse outer halo.  This suggests that the radiative transfer of \lya\ photons in high-redshift LAEs is complicated, with the interstellar-medium geometry and/or outflows playing a significant role in galaxies at these redshifts.
\end{abstract}


\section{Introduction}

High-redshift \lya\ emitting galaxies (LAEs) are some of the most intriguing objects in the distant universe.  Their strong \lya\ emission was thought to be indicative of the first galaxies \citep{partridge67}, implying that they could possibly contain the first stars and likely be composed of pristine gas.  However, recent studies of their physical properties imply that some of these galaxies may be more evolved, with many LAEs exhibiting rest-frame ultraviolet colors indicative of modest-to-moderate dust extinction \citep[e.g.,][]{pirzkal07,lai07,finkelstein08,finkelstein09a,pentericci09,ono10}.  As \lya\ photons are resonantly scattered by neutral hydrogen, galaxies with dust would be unlikely to exhibit \lya\ in emission.  Thus, just how \lya\ escapes from a galaxy with a dusty interstellar medium is an outstanding question in the study of distant galaxies.

As LAEs have been selected on the basis of their \lya\ emission, some mechanism must allow the escape of these photons.  One possibility is that the \lya\ photons we see have been shifted out of resonance by scattering off of the receding edge of an outflow in the ISM.  Evidence for outflows has been observed many times in the typically more evolved Lyman break galaxies \citep[LBGs; e.g.,][]{shapley03,bielby10} as a velocity difference between \lya\ emission and ISM absorption features.  Only recently have outflows been shown to exist in LAEs, as \citet{mclinden10} discovered that in two LAEs at $z \sim$ 3.1 \lya\ emission had a slightly higher redshift than the rest-frame optical [O\,{\sc iii}] emission, which is thought to come from H\,{\sc ii} regions at the systemic redshift.  In either case, much of the \lya\ emission is shifted redward of the resonance line at 1216 \AA, and thus will have an easier chance of escaping, even in a uniform ISM.

Alternatively, if the line emission is primarily at resonance, much of the \lya\ emission can still escape if the ISM is primarily clumpy, as the \lya\ photons will scatter off of the clumps, and be screened from seeing much of the dust \citep{neufeld91,hansen06}.  This type of ISM geometry can explain the dustiness of LAEs at z $\sim$ 4.5 \citep {finkelstein08,finkelstein09a}, and can also explain the large number of high \lya\ equivalent widths (EWs) which have been observed \citep[e.g.,][]{kudritzki00,malhotra02,finkelstein07}.

In either of these radiative-transfer scenarios, any detected \lya\ emission will be spatially de-correlated from its origination point within its host galaxy.  By comparing \lya\ emission from a narrowband filter to the rest-frame UV emission from a neighboring broadband filter, one can diagnose whether this is the case; if \lya\ has undergone any extreme radiative-transfer effects, this should reveal itself in a larger size in the \lya\ emission, as well as possibly a diffuse \lya\ halo.  

However, all currently known LAEs have been discovered via ground-based narrowband imaging, which even in the best seeing conditions cannot resolve the extremely small physical sizes of LAEs of 1 -- 2 kpc \citep[][Malhotra, S.\ et al.\ 2010, in prep]{bond09}.  Here we report on the results of a new {\it Hubble Space Telescope} imaging survey designed to search for resolved \lya\ emission from LAEs at $z \sim$ 4.4 using the F658N narrowband filter on the Advanced Camera for Surveys.  Throughout we use the AB magnitude system, where m$_{AB}$ = $-$2.5 log (f$_{\nu}$) $-$ 48.6 mag.  Where applicable, we assume a concordance cosmology, with H$_\mathrm{o}$ = 70 km s$^{-1}$ Mpc$^{-1}$, $\Omega_{m} = 0.3$ and $\Omega_{\Lambda} = 0.7$.  At $z =$ 4.4, this corresponds to an angular scale of 6.671 kpc arcsec$^{-1}$. 

\section{Data}

\subsection{Observations}
Thanks to the successful repair of the Advanced Camera for Surveys (ACS) during Servicing Mission 4 (SM4) to the {\it Hubble Space Telescope} ({\it HST}), we were able to obtain ACS parallel imaging during the Early Release Science \citep[ERS;][]{windhorst10} Wide Field Camera 3 (WFC3) observations of the Great Observatories Origins Deep Survey (GOODS) Chandra Deep Fields -- South (CDF -- S).  We obtained 11 orbits per pointing over 8 independent pointings.  Due to the location of the WFC3 fields, all of the ACS pointings overlapped the GOODS CDF-S field, which has existing deep public data in the F435W, F606W, F775W and F850LP ACS filters (as well as a wealth of other multiwavelength data).  A detailed summary of the primary WFC3 ERS images, as well as their layout and analysis is given by \citet{windhorst10}.

\begin{figure}[t]
\epsscale{1.1}
\plotone{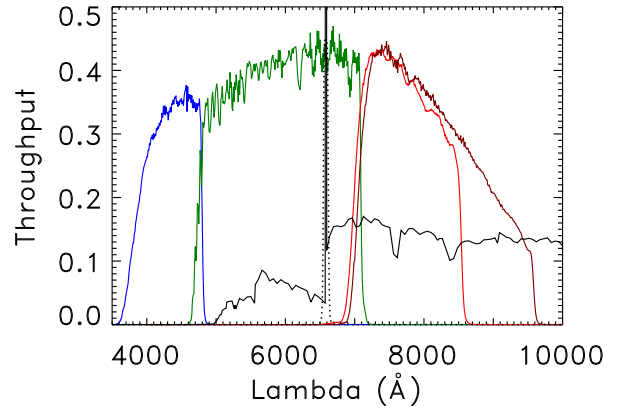}
\caption{A model spectrum of a $z$ = 4.42 LAE, showing the {\it HST} ACS F435W, F606W, F775W and F814W bandpasses in blue, green, red and brown, as well as the F658N bandpass (dotted line).  From z = 4.38 -- 4.45, \lya\ passes through the F658N bandpass, allowing imaging of \lya\ light at high redshift.}\label{fig:filters}
\end{figure}

We split each parallel pointing into 9 orbits with the F658N narrowband filter and 2 orbits with the F814W broadband filter.  With a central wavelength of 6584 \AA~and a full-width at half-maximum (FWHM) of 73 \AA, the F658N observations will detect \lya~(which has $\lambda_{rest}$ = 1215.67 \AA) from redshifts 4.386 $\leq z \leq$ 4.445.  A model spectrum of a LAE at $z$ = 4.42 is shown in \fig{fig:filters}.  At z = 4.4, the existing GOODS ACS data cover rest-frame wavelengths of $\sim$ 800 \AA\ (F435W), 1100 \AA\ (F606W), 1400 \AA\ (F775W) and 1600 \AA\ (F850LP).  The F814W data will cover the continuum at $\sim$ 1500 \AA, providing an independent observation in addition to the existing GOODS dataset.

\subsection{Data Reduction}
The raw ACS data were downloaded from the Space Telescope Science Institute (STScI) archive.  The ACS data were taken in 31 separate visits, with typical dithers within each visit of $<$ 20\arcs.  Images from visits with central pointings separated by less than 25\arcs\ were reduced together, yielding 17 separate reductions.  The raw images were processed using the {\tt calacs} task, which is in the stsdas package in IRAF\footnote[1]{IRAF is distributed by the National Optical Astronomy Observatory (NOAO), which is operated by the Association of Universities for Research in Astronomy, Inc.\ (AURA) under cooperative agreement with the National Science Foundation.}.  This task provides routine calibration, including bias, dark and flat-field corrections, using the most recent ACS reference files taken after SM4 retrieved from the {\it HST} archive.  ACS data obtained after SM4 suffer a low-level striping pattern.  We implemented a custom-built script (provided by NAG) to remove this pattern prior to the flat-field correction.

\begin{figure*}
\epsscale{1.0}
\vspace{6mm}
\plotone{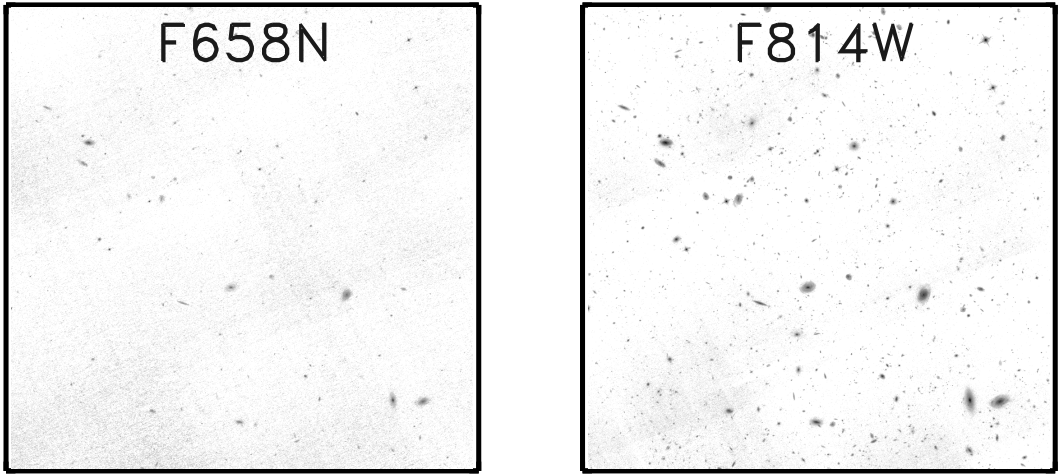}
\caption{F658N (left) and F814W (right) image of GOODS-S section 23.  This was the only GOODS-S section which was completely covered by our observations.  The narrow bandpass of F658N is apparent when comparing the depths of these images, as many fewer objects are apparent in F658N even though the exposure times were longer.}\label{fig:images_bin}
\end{figure*}

The calibrated and pattern-corrected images were cleaned of cosmic-rays, distortion-corrected, registered and combined using the task {\tt multidrizzle} \citep{koekemoer02}. Upon completion of the initial run of {\tt multidrizzle} in each visit, it was apparent that the registration was not ideal, as stars in the combined images appeared elongated.  We thus ran custom-built scripts (provided by AMK) on a visit-by-visit basis to correct the World Coordinate System (WCS) in the headers of the individual frames, solving for the relative astrometric shifts between frames \citep{windhorst10}.  {\tt Multidrizzle} was then run a second time to create a final, combined image for each visit.  As we planned to use the existing GOODS ACS data in our analysis, we used the GOODS ACS image sections as reference images when running {\tt multidrizzle}.  In order to correct for small astrometric differences between the GOODS data and these new ACS data, we first ran {\tt multidrizzle} with no reference image, and performed photometry using the source extractor software package (hereafter SExtractor; \citet{bertin96}) to identify objects in the image.  We did the same to the relevant GOODS section, and then ran the IRAF tasks {\tt xyxymatch} and {\tt geomap} to match common objects between the two frames, and construct a shift file.  This shift file was used for the final iteration of {\tt multidrizzle} to create F658N and F814W images matched to each GOODS section covered by our observations.  Our final dataset was composed of one image in each of the two filters for the 11 GOODS-S sections that we covered: 12, 13, 14, 22, 23, 24, 32, 33, 34, 42 and 43.  Typical exposure times in the reduced datasets are 11000 s in F658N and 2200-2500 s in F814W.  Images of section 23 in the F658N and F814W bands are shown in \fig{fig:images_bin}.

\subsection{Catalog Construction}
We created narrowband-selected catalogs for each observed GOODS-S section using SExtractor in two image mode, with the F658N image for each section as the detection image, and our F658N and F814W images, as well as the GOODS F435W, F606W, F775W and F850LP images as the measurement images.  We used identical SExtractror parameters as used in GOODS.  The final catalog encompassing all covered sections includes 3081 narrowband-selected objects, with fluxes measured in 0.7\arcs\ diameter apertures, as well as estimates of the total flux using SExtractor's MAG$\_$AUTO measurement.

\begin{figure}[t]
\epsscale{1.1}
\plotone{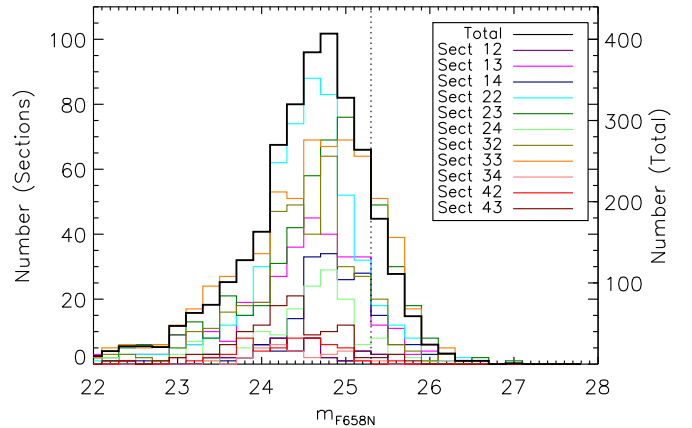}
\caption{Number counts of objects in the F658N data.  The colored histograms denote the numbers in each section, with the values being given by the left-hand vertical axis.  This thick black line denotes the total combined number counts, with the values being given by the right-hand vertical axis.  The varying number of objects per section is related to the amount of section area which received F658N coverage.  The dotted line denotes where the total number counts fall to 50\% of their peak value, which is at m$_{F658N}$ = 25.3.}\label{fig:number}
\end{figure}

Initial flux errors were taken to be the calculated SExtractor errors.  We checked these errors by measuring our own errors in each image.  This was done by measuring the flux in 10$^{4}$ randomly placed 0.7\arcs-diameter apertures in each of the six images, and then examining the spread of these fluxes (this was done in GOODS-S section 23, as this was the only section which had complete coverage by our F658N and F814W data due to the unfavorable positioning of the parallel exposures with respect to the GOODS sections).  The characteristic 1 $\sigma$ error for each image was taken as the $\sigma$ of a Gaussian fit to a histogram of the flux distribution.  Comparing this error to the median SExtractor error in each image, we find that SExtractor underestimated the errors by up to $\sim$ 20\% (with the exception of the F814W data, where SExtractor overestimated the errors by 18\%).  While we trust our independently computed errors as being indicative of the global uncertainty in the image, the errors computed by SExtractor include information on the local background.  We thus scaled the median SExtractor error to match the global uncertainty in each band.  The derived 5 $\sigma$ limits for each band in a 0.7\arcs-diameter aperture are: 25.0 (F658N), 27.1 (F814W), 27.4 (F435W), 27.6 (F606W), 27.0 (F775W) and 26.8 (F850LP).  Number counts of objects in the F658N images are shown in \fig{fig:number}, showing a peak at m$_{F658N}$ $\sim$ 24.8 mag.

\begin{figure*}
\epsscale{0.9}
\vspace{6mm}
\plotone{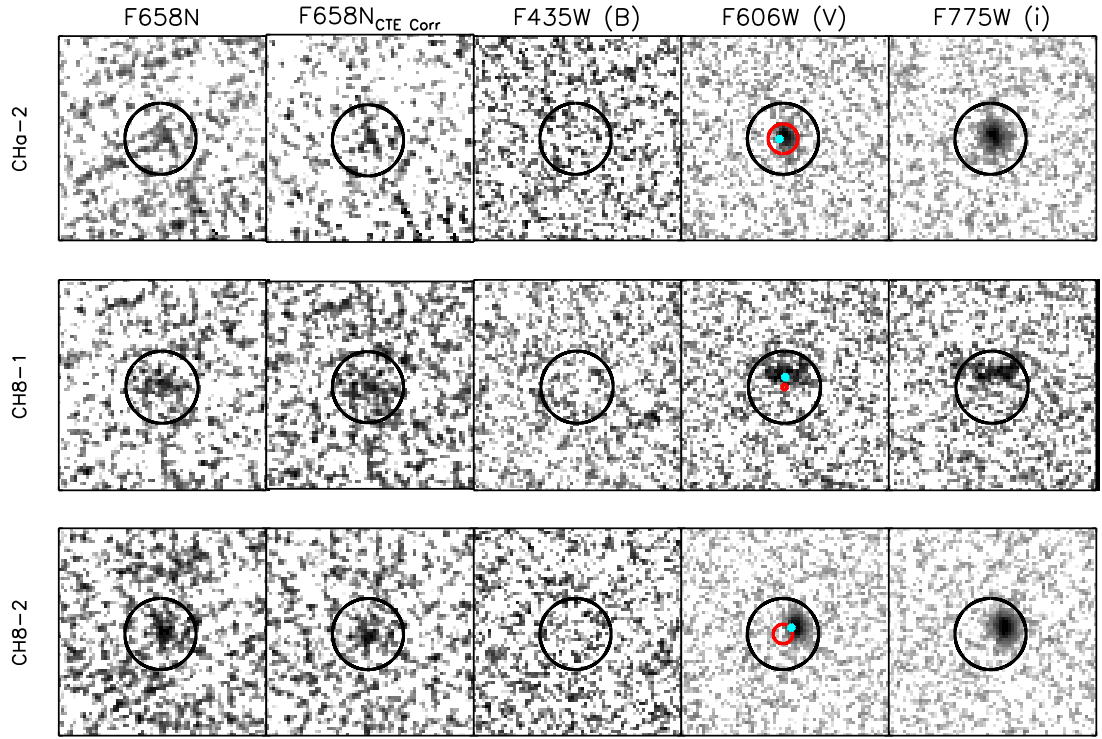}
\caption{Cutouts of the three LAEs in our sample, 2\arcs\ on each side.  The black circles are centered on the F658N centroid, with a 0.7\arcs\ diameter.  The F658N data is from our program, while the remaining images are from the GOODS dataset.  Object CHa-2 is near an image edge, thus the lower-right corner of this stamp is not real data.  Although trailing due to charge transfer inefficiency is apparent in object CHa-2 (as blurring from lower-left to upper-right), the narrowband flux is still detected at 3.9 $\sigma$ in a 0.7\arcs~diameter aperture.  The second column shows the CTE-corrected stamps, as discussed in \S 5.2.4.  The CTE-correction needs to be verified with newer calibration data before it can be folded into the analysis, but improvement in the background can be seen, especially near LAE CHa-2.  The F435W data probes rest-frame $\lambda$ $\sim$ 800 \AA\ at $z = $4.4, thus the non-detections in this image are expected.  The cyan circle in the F606W images denotes the \lya\ emission centroid after correction for a systematic positional offset between the WCS of the new F658N data and the existing GOODS data.  The red circle denotes the 1 $\sigma$ uncertainty on the positional offset.  These uncertainties are large in CHa-2 and CH8-2.  However, the offset can be computed to a high precision in CH8-1, and we find that the apparent offset between the \lya\ and UV emission can be explained by WCS differences in the two datasets.}\label{fig:thesislaes}
\end{figure*}

\begin{deluxetable*}{cccccccccc}
\tablecaption{Properties of the LAEs}
\tablewidth{0pt}
\tablehead{
\colhead{Object} & \colhead{Redshift} & \colhead{RA} & \colhead{Dec} & \colhead{m$_{F658N}$} & \colhead{m$_{F606W}$} & \colhead{m$_{F775W}$} & \colhead{r$_{h,F658N}^{APER}$} & \colhead{r$_{h,F775W}^{APER}$} & \colhead{Rest EW}\\
\colhead{$ $} & \colhead{$ $} & \colhead{(J2000)} & \colhead{(J2000)} & \colhead{(mag)} & \colhead{(mag)} & \colhead{(mag)} & \colhead{(kpc)} & \colhead{(kpc)} & \colhead{(\AA)}
}
\startdata
CHa-2&4.414&03:32:39.77&-27:51:14.97&24.90 $\pm$ 0.28&26.57 $\pm$ 0.09&25.56 $\pm$ 0.06&0.73 $\pm$ 0.13&0.67 $\pm$ 0.05&167\\
CH8-1&4.434&03:32:49.01&-27:49:02.08&25.20 $\pm$ 0.22&27.06 $\pm$ 0.14&26.54 $\pm$ 0.14&1.40 $\pm$ 0.16&1.07 $\pm$ 0.11&176\\
CH8-2&4.433&03:32:54.04&-27:50:00.83&24.64 $\pm$ 0.20&25.93 $\pm$ 0.05&24.98 $\pm$ 0.04&1.33 $\pm$ 0.18&0.67 $\pm$ 0.03&53\\
\enddata
\tablecomments{The position is the centroid of the F658N counterpart.  All magnitudes were measured with 0.70\arcs~diameter apertures.  Half-light radii (r$_{h}$) were converted from arcsec to kpc assuming all objects are at $z = 4.4$, which gives an angular scale of 6.671 kpc arcsec$^{-1}$ for our assumed cosmology.  The rest-frame equivalent widths are from \citet{finkelstein09a}, and are based on the ground-based narrowband imaging.}
\end{deluxetable*}

\section{Sample Selection}

\subsection{Spectroscopically Confirmed LAEs}
In \citet{finkelstein08} and \citet{finkelstein09a}, a sample of 14 LAEs were discovered in the GOODS-S field using ground-based narrowband selection.  These studies used three overlapping narrowband filters, centered at 6560 (hereafter NB656), 6650 (NB665) and 6730 (NB673) \AA\ to discover LAEs at $z \approx$ 4.4 -- 4.5.  Samples of 4, 2 and 8 candidate LAEs were discovered in the three images, with 5 $\sigma$ depths of 24.9, 25.0 and 25.2 mag, respectively.  The ACS F658N filter can measure \lya~emission from galaxies at z = 4.38 -- 4.45; thus it would also observe \lya\ from objects discovered in the red half of the NB656 filter, or in the blue half of the NB665 filter.  Of the six candidate LAEs discovered in these two filters, three fall in the area covered by our F658N observations.  These three objects are CHa-2, CH8-1 and CH8-2, using the nomenclature from \citet{finkelstein09a} (where CHa denotes CDFS H$\alpha$, i.e., NB656, and CH8 denotes CDFS H$\alpha$+80 \AA, i.e. NB665).  Images of these three objects are shown in \fig{fig:thesislaes}.

\begin{figure}[t]
\epsscale{1.1}
\plotone{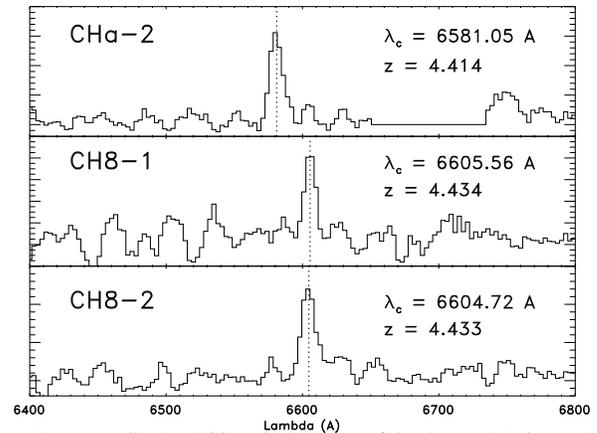}
\caption{Magellan/IMACS optical spectra of the three LAEs in our data, centered around the observed \lya\ emission line.  The vertical scale is in arbitrary flux units.  From the position of the emission line, and the lack of any other emission line in the full spectra, we confirm that all three objects are LAEs at z $\approx$ 4.4.  This confirms that the flux we detect in the F658N image is \lya\ emission from these objects.}\label{fig:spectra}
\end{figure}

Although these objects were previously selected via narrowband observations from the ground, they had yet to be spectroscopically confirmed.  We recently obtained optical spectroscopy of these three objects with the Inamori Magellan Areal Camera and Spectrograph (IMACS) at the Magellan Baade Telescope on 11-12 November 2009 (NOAO PID 2009B-0371, PI Finkelstein).  The full details of this spectroscopic dataset will be presented in a future paper (Zheng et al.\ in prep), but in brief, each object was observed as part of a 4-hour slit-mask integration, with the f/2 camera and the 300 lines/mm grating blazed at 17.5$^\mathrm{o}$ (giving R $\simeq$ 1000).  The reduced, one-dimensional spectra of these three objects are shown in \fig{fig:spectra}.  Each object exhibits a single emission line with no significant continuum light, indicative of Ly$\alpha$ emission at high redshift.  Fitting a Gaussian curve to these emission lines, we find redshifts of CHa-2, CH8-1 and CH8-2 of 4.414, 4.434 and 4.433, respectively, placing the \lya\ emission line of each object in the bandpass of the ACS F658N filter.
 
Examining these objects  in \fig{fig:thesislaes}, they all appear robustly detected in the F658N image.  However, especially in the case of CHa-2, the noise due to poor charge transfer efficiency (CTE) in the nearly decade-old CCDs onboard ACS is apparent.  Nonetheless, when we consult our narrowband selected catalog, we find that CHa-2, CH8-1 and CH8-2 are all formally detected, with detection significances of 3.9, 5.0 and 5.4 $\sigma$, respectively.  Combined with the fact that LAEs were previously known to reside at these locations, we are confident that we are in fact detecting \lya\ emission with ACS.

\subsection{Photometric Redshift Selection}
In addition to objects previously selected on the basis of their \lya\ emission at $z \approx$ 4.4, we have also examined the F658N images for objects which are likely to reside at $z \sim$ 4.4 based on their spectral energy distribution.  We selected objects at this redshift from two catalogs, both from S. Cohen et al.\ (2010, in prep).  The first consists of $\sim$ 15000 objects with spectro-photometric redshifts computed using both ACS broadband and grism slitless spectroscopic data from the Probing Evolution and Reionization Spectroscopically (PEARS) program (PI S. Malhotra).  The second catalog consists of $\sim$ 8000 photometric redshifts measured over the entire GOODS-S region, using VLT/VIMOS U \citep{nonino09}, GOODS/ACS v2.0 B, V, i$^{\prime}$, z$^{\prime}$, and GOODS VLT/ISAAC v2.0 J, H and K-band data \citep{retzlaff10}.

We examined these objects for galaxies with best-fit (spectro-)photometric redshifts of 4.38 $\leq$ z $\leq$ 4.45, placing any \lya\ emission in the F658N bandpass.  We also included objects that had this redshift slice contained within the 68\% confidence range on their spectro-photometric redshift.  We found 106 objects meeting these criteria.  We then matched these objects to our F658N catalog, using a matching radius of 0.5\arcs, and we found six objects that have F658N counterparts.  The low number of matched objects is expected, as only galaxies exhibiting \lya\ emission at the specific redshift placing it in the F658N bandpass would be detected in the narrowband data.  These objects were visually inspected in the F658N data.  Of these six objects, only two have moderate narrowband excesses (m$_{F606W}$ - m$_{F658N}$ = 0.6 and 1.8 mag).  However, both objects have significant detections in the F435W-band.  This band is entirely blueward of both the Ly$\alpha$ and Lyman continuum break at z $\approx$ 4.4 (see \fig{fig:filters}); thus there should not be a F435W detection if these objects were truly at $z \approx$ 4.4.  We conclude that these two objects are low-redshift interlopers, and we exclude them from further study.

\section{Results}

With our sample of three F658N-detected high-redshift LAEs, we investigate their light profiles, as well as the location of their \lya\ emission.

\subsection{Positional Differences Between \lya\ and Rest-Frame UV Emission}
In a number of objects the \lya\ emission appears offset from the centroid of the rest-frame UV emission.  If this effect is real, it is quite interesting, as it could indicate that \lya\ is escaping only after scattering off of gas or dust outside the primary stellar population, perhaps due to outflows in the ISM \citep[e.g.,][]{windhorst98, waddington99}.  However, we first need to investigate if the offset is real, or if it is an artifact of mismatches between the WCS of the new F658N data and the existing GOODS-S data.

We investigated these offsets by examining the relative pixel positions of all objects in the images around the LAEs.  To find these objects, we first ran SExtractor on both the F658N and F606W images, using each image as its own detection image, such that we obtained object coordinates native to each image.  On a LAE by LAE basis, we first searched the F658N catalog for all objects in a given section, excluding objects near the edge of our images, as well as objects below the point where the number counts fall to 50\% of their peak value, which is at 25.3 mag.  We then computed the distance in pixels from the LAE to each of these objects.  We selected objects within a threshold radius, which ranged from 500 -- 3000 pixels in 100 pixel increments, and matched them to objects in the F606W catalog, keeping objects that were matched within 20 pixels (which is larger than the largest apparent shift; see \fig{fig:thesislaes}).  By including only objects near the LAE, we ensure that we are locally measuring any offset between the F658N and F606W image frames.  At a 1000 pixel radius, on average a dozen matches were found, increasing to $\sim$ 40 matches by 2000 pixels.  The pixel offsets were then computed as the mean difference between the narrowband position and the broadband position for each of the matched objects.    An estimate of the uncertainty on these shifts was taken to be the standard deviation of the positional differences for the matched objects.

\begin{figure}[t]
\epsscale{1.2}
\plotone{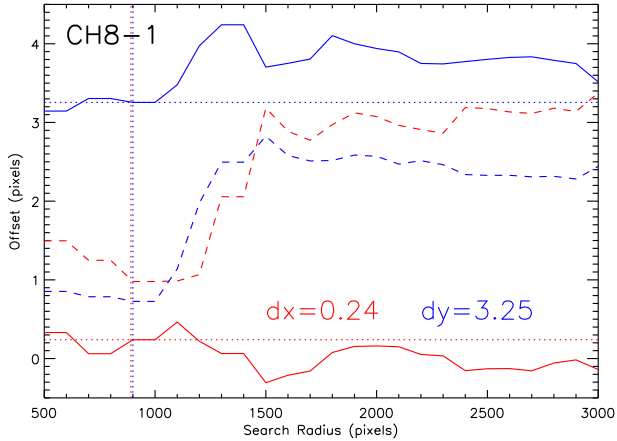}
\caption{The derived pixel offsets between the F658N data and the GOODS F606W data for objects near LAE CH8-1, versus the search radius used to find objects to derive the offset.  Red and blue denote the x and y offsets, respectively, while the solid and dashed lines denote the offset and offset uncertainty.  Offsets were derived using objects within 500 -- 3000 pixels, in 100 pixel increments.  The value of the offset was defined to be the offset derived from the search radius which produced the smallest offset uncertainty (designated by the dotted lines).  In this object, the offset uncertainties for both x and y reached a minimum at 900 pixels, thus the pixel offsets were taken to be the values at that search radius.  At larger search radii, the offset uncertainties increase dramatically.}\label{fig:offset}
\end{figure}

\fig{fig:offset} shows an example of this process, showing the results for LAE CH8-1.  We plot lines showing both the pixel offsets, as well as the offset uncertainties as a function of search radius.  We chose offset values for each object to be the pixel offset value at the radius where the offset uncertainty was a minimum.  For this object, the x-offset is negligible (0.24 $\pm$ 0.98 pixels), while the y-offset is significant, at 3.25 $\pm$ 0.73 pixels.  Investigating \fig{fig:thesislaes}, this offset would move the broadband counterpart down vertically, bringing it more in line with the narrowband position.  This is shown by the cyan circle in \fig{fig:thesislaes}.  A similar process was done for the remaining objects, and these offsets are tabulated in Table 2.

\begin{figure*}[!ht]
\epsscale{1.0}
\plotone{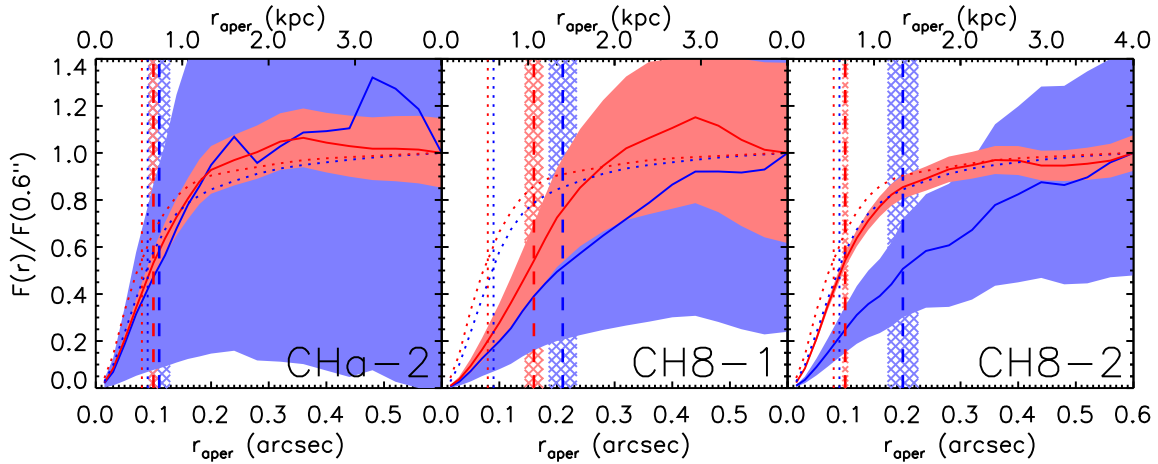}
\caption{Measured curves-of-growth from aperture photometry for the objects in our sample in blue and red for F658N and F775W, respectively.  The shaded regions show the 1 $\sigma$ uncertainties on the CoGs.  The colored dashed lines denote the half-light radii, with the cross-hatched regions denoting the 1 $\sigma$ uncertainties on the radii.  The colored dotted lines denote the resolution limit of the given image derived from the sizes of stars in the images, while the dotted curves show the CoGs of the image PSFs.  The CoGs of \lya\ appear more extended than those of the rest-frame UV continuum, resulting in larger \lya\ half-light radii.}\label{fig:profile}
\end{figure*}

In \fig{fig:thesislaes} we show the corrected \lya\ emission position by a small cyan circle, and the typical offset error as a red circle (where the radius of the circle is the mean of the x- and y-offset errors for a given object).  We find that in CHa-2 and CH8-2, the uncertainties on the derived pixel offsets are large, and thus any apparent offset would be at low significance.  However, in CH8-1, the offset uncertainties are small, and we can see that the computed offset is consistent with the centroid of the UV emission.  Thus, while the \lya\ and UV emission in CH8-1 appear to be offset, this can be explained by relative offsets between the two datasets. 

We conclude that while it is possible that ISM scattering effects can result in an offset between the apparent positions of \lya\ and rest-frame UV emission, we cannot conclusively support this with our data.  The most convincing offset is in CH8-1, as the offset is large.  However, this is also the only object where the offset uncertainties are small enough that we can reasonably correct the \lya\ emission position, and we find that its corrected position is then coincident with the UV emission.  In addition, when inspecting the F814W data taken at the same time as the F658N data, the apparent offset between F658N and F814W is much less.  We move forward assuming that the \lya\ emission is coincident with the rest-frame UV emission in all objects.

\subsection{Physical Size of \lya\ Emission}

\subsubsection{Individual Objects}

In order to measure the physical sizes of the LAEs in our sample in both their \lya\ and rest-frame UV continuum light, we employed the method of \citet{bond09}.  We first cut out 101$\times$101 pixel (3\arcs$\times$3\arcs) postage stamp FITS images centered on each LAE in both the F658N and F775W data (we used the F775W rather than the F606W data for the rest-frame UV as it is completely redward of \lya\ at this redshift).  We then ran SExtractor on each stamp, using the stamp as both the detection and measurement image to determine the flux-weighted center of the object, as well as to determine whether an object is made up of sub-clumps that might have been split up by SExtractor.  Experimenting with various values of the DEBLEND\_NTHRESH parameter, we found that none of our LAEs can be split into multiple objects; thus we conclude that all objects in our sample are composed of single dominant components.  Inspecting the SExtractor results, we find that (as expected) each object is detected in both F658N and F775W.  

\begin{deluxetable}{ccc}
\tablecaption{Corrections to Narrowband Emission Position}
\tablewidth{0pt}
\tablehead{
\colhead{Object} & \colhead{$\Delta$x} & \colhead{$\Delta$y}\\
\colhead{$ $} & \colhead{(pixels)} & \colhead{(pixels)}
}
\startdata
CHa-2&$-$1.17 $\pm$ 5.10&$-$0.08 $\pm$ 4.66\\
CH8-1&\phantom{$-$}0.24 $\pm$ 0.98&\phantom{$-$}3.25 $\pm$ 0.73\\
CH8-2&\phantom{$-$}2.74 $\pm$ 4.16&\phantom{$-$}2.09 $\pm$ 2.28\\
\enddata
\tablecomments{The derived pixel corrections to the F658N emission
  position due to WCS differences between the F658N and the GOODS
  data.  The corrected narrowband emission position is given by the
  cyan circles in \fig{fig:thesislaes}, while the positional
  uncertainties are shown by the red circles..}
\end{deluxetable}

Using the SExtractor-derived center in the F658N and F775W images, respectively, we measured the flux in a series of 32 apertures, with radii ranging from 0.015 -- 1.2\arcs\ using SExtractor.  In order to ensure that the CTE-affected background was subtracted as well as possible, we manually subtracted the background prior to running SExtractor, using the iterative mean computed with the IDL task djs\_iterstat.pro.  We then forced SExtractor to assume a background value of zero.  Previously measured half-light diameters of LAEs are $\sim$ 0.2 -- 0.4\arcs\ \citet{bond09, bond10}, thus we assume that the flux at a radius of 0.6\arcs\ approximates the total flux.  We then compute the radius at which the flux is half of the flux at $r = 0.6$\arcs, and use that as an estimate of the half-light radius (r$_{h}$).  

Values of r$_{h}$ were computed for each object in F658N and in F775W, and are tabulated in Table 1.  The curves-of-growth (CoGs) of each object in both bands are shown in \fig{fig:profile}.  We also show the uncertainty in the CoGs as the shaded region, using a similar exercise as explained in \S 2.3 to compute the flux uncertainty in all 32 apertures (using the flux uncertainty from the appropriate image section).  We then used these errors to compute uncertainties on our derived half-light radii by running a series of 10$^{4}$ Monte Carlo simulations.  In each Monte Carlo simulation, we vary the flux at each point in the CoG by a random number (drawn from a Gaussian distribution centered at zero with $\sigma$ = 1) multiplied by the flux uncertainty, and rederive the half-light radius.  The uncertainty on the radius is then the standard deviation of the radii from the simulations.  The radii uncertainties are shown as cross-hatched regions in \fig{fig:profile}.  Typical uncertainties on r$_{h}$ are $\sim$ 0.02\arcs\ in F658N and $\sim$ 0.01\arcs\ in F775W.

To determine whether a particular object is resolved, we performed the above analysis on a point-spread-function (PSF) made from stars in both the F658N and F775W data.  These PSFs were made by adding together images of five stars identified in section 23.  We first cut out 101$\times$101 pixel postage stamps around each star.  In order to be sure the stars were centered, we computed the difference between the centroid of the star and the center of the array.  If the difference was more than 0.2 pixels in either direction, we subsampled the image by a factor of 10, and shifted the star by one pixel for each tenth of a pixel it was offset from the center (the subsampling was done using the IDL function {\tt frebin}, which uses bilinear interpolation).  The image was then binned back down to the native resolution.  This process was run iteratively on each star until they were all $<$ 0.2 pixels from the array center.  Each star was normalized to its peak flux.  The PSF was then calculated as the median of the five stars at each pixel position, and then normalized to a total flux of 1.  Measuring the half-light radii of the PSF in each band in the same manner as above, we measure an image resolution of r$_{h}$ = 0.09\arcs\ in F658N, and r$_{h}$ = 0.08\arcs\ in F775W.  Objects with r$_{h}$ at or less than these values are considered unresolved at the limit of {\it HST}$+$ACS in their respective bands.

For our sample of LAEs, we found half-light radii in the F658N image of 0.11\arcs\ $\pm$ 0.02\arcs, 0.21\arcs\ $\pm$ 0.02\arcs\ and 0.20\arcs\ $\pm$ 0.02\arcs\ for CHa-2, CH8-1 and CH8-2, respectively.  In the F775W image, we found LAE half-light radii of 0.10\arcs\ $\pm$ 0.01\arcs, 0.16\arcs\ $\pm$ 0.01\arcs\ and 0.10\arcs\ $\pm$ 0.01\arcs\ for CHa-2, CH8-1 and CH8-2, respectively.  Comparing these sizes to the PSFs discussed above, we find that we can definitively resolve 2/3 LAEs in the F658N image, and 1/3 LAEs in the F775W image.  At $z = 4.4$, the angular scale is $\sim$ 6.671 kpc arcsec$^{-1}$ (for our assumed cosmology); thus these sizes correspond to 0.7 -- 1.4 kpc in \lya, and 0.7 -- 1.1 kpc in the rest-frame UV continuum\footnote[2]{The Year 7 Wilkinson Microwave Anisotropy Probe cosmology (H$_\mathrm{o}$ = 70 km s$^{-1}$ Mpc$^{-1}$, $\Omega_{m} = 0.27$ and $\Omega_{\Lambda} = 0.73$ \citep{komatsu10}) gives an angular scale of 6.899 kpc arcsec$^{-1}$, which would give physical sizes 4.3\% larger than our assumed cosomology.}.

\subsubsection{Stacking Analysis}
As shown in Table 1, the signal-to-noise of the individual detections in the F658N image are not large.  Thus, in order to obtain a more robust estimate of the {\it average} half-light radii of LAEs, we have performed a stacking analysis.  \citet{hathi08b} show in detail how such image stacking is justified for similar galaxies at similar redshifts, using the HUDF B, V and i$^{\prime}$ dropouts at $z =$ 4, 5 and 6, respectively.

Using the cutout stamps described in the above section, we first centered each LAE on the central pixel of each stamp using the iterative technique described above for the PSF, requiring the SExtractor-derived center to be within 0.2 pixels of the center of the stamp.  This step was performed separately for each object for each band, such that the F658N stamps were centered on the F658N emission, and the F775W stamps were centered on the F775W emission.  Each centered LAE stamp was then normalized to its peak flux.  A stacked image was then created in each band by taking the median of each pixel value from all three LAEs.  \fig{fig:stack} shows the stacks of the three LAEs in both bands, with the contours denoting levels of constant brightness, as well as three-dimensional surface brightness profiles.

We measured half-light radii of each of the two stacks (one for each band) in the same manner as the above section.  The results from this analysis are shown in \fig{fig:flux_stack}.  These stacking results confirm our observations of the individual objects, in that the \lya\ emission is more extended than the rest-frame UV continuum emission, with r$_{h}$ = 0.16 $\pm$ 0.01\arcs\ in F658N, and r$_{h}$ = 0.10 $\pm$ 0.01\arcs\ in F775W.  These angular sizes correspond to physical half-light radii of 1.07 $\pm$ 0.08 and 0.67 $\pm$ 0.05 kpc for the F658N and F775W emission, respectively.

\section{Discussion}

\subsection{Rest-Frame UV Emission}

Inspecting \fig{fig:profile}, one can see that LAEs are compact in their rest-frame UV continuum with half-light radii of r$_{h}$ $<$ 1.1 kpc in all three objects.  This is consistent with previous studies of high-redshift galaxies.  \citet{ferguson04} studied the rest-frame UV sizes of Lyman-break-selected galaxies (LBGs) at $z >$ 3, and photometric-redshift-selected galaxies at 1 $< z <$ 3 using data from {\it HST}.  They found half-light radii from 0.25 -- 0.4\arcs\ at $z > 2$ ($\sim$ 2 -- 3 kpc), rising to r$_{h}$ $\sim$ 0.65\arcs\ at $z \sim$ 1 ($\sim$ 5 kpc) for galaxies with 0.7$L^{\ast} < L_{UV} < 5L^{\ast}$.  Similar size evolution has been found to extend out to z $\sim$ 6 \citep{hathi08} and $z =$ 7 -- 8 \citep{oesch10b}, where $L_{UV} \sim L^{\ast}$ LBGs have r$_{h}$ $\sim$ 1 kpc.

Relatively few LAEs have had their morphologies studied.  Recently, \citet{bond09} studied the rest-frame UV morphologies of a sample of LAEs at z $\sim$ 3.1 from the MUSYC survey \citep{gawiser06b}.  They found that LAEs are typically at least as compact as LBGs, with r$_{h}$ $\lesssim$ 2 kpc, and that the \lya\ emission is likely coincident with the UV emission (within $<$ 1 kpc).  \citet{gronwall10} studied the same sample, examining the better detected LAEs (S/N $>$ 30) in greater detail, finding that their rest-frame UV light is very concentrated, and that they have Sersic indices indicative of disk-like morphologies in most instances (0 $<$ n $<$ 2).  

\subsection{\lya\ Emission}

\subsubsection{Previous Results}
 Prior to this study, only two high-redshift \lya-selected galaxies have been imaged in their \lya\ light at {\it HST} resolution (i.e., using space-based narrowband data), published recently by \citet{bond10}.  In this study, \citet{bond10} obtained {\it HST}/WFPC2 F502N imaging of $z \sim$ 3.1 LAEs, obtaining detections of two out of the eight LAEs they targeted.  They concluded that these objects have \lya\ half-light radii $<$ 1.5 kpc, similar to their rest-frame UV sizes, with the \lya\ emission coincident within 0.5 kpc of the rest-frame UV emission.  \citet{rhoads09} also examined the relative sizes of LAEs in \lya\ and the UV continuum using ACS grism spectroscopic data from the PEARS survey by examining the sizes of the objects in the spatial dimension.  They did not find evidence of an extended \lya\ halo in a stack of the spectra from all 39 z $\sim$ 5 galaxies in their sample.  However, when stacking only the 10 galaxies with \lya\ observed in emission, they found that the spatial width of the spectrum at the position of \lya\ had FWHM = 0.26\arcs, while the same measurement on the adjacent UV continuum yielded FWHM = 0.19\arcs, suggesting possible extended \lya\ halos in these objects.

Our positive detections of the three spectroscopically confirmed LAEs more than doubles the total number of high-redshift LAEs with high-resolution imaging of their \lya\ light.  Investigating the \lya\ light profiles of our LAEs, we find that the \lya\ emission appears relatively compact as well, with the half-light radius in every object at $\leq$ 1.4 kpc, and the mean size of $\sim$ 1.2 kpc consistent with the \lya\ sizes of the two galaxies studied by \citet{bond10}.

\subsubsection{Individual Objects}
Comparing the CoGs of the \lya\ and rest-frame UV continuum light in individual LAEs in \fig{fig:profile}, we find that CH8-1 and CH8-2 have \lya\ half-light radii {\it larger} than the rest-frame UV, while CHa-2 is near the limit of our resolution in both the \lya\ and UV continuum light.  The CoGs of CH8-1 and CH8-2 are very similar, with the rest-frame UV (F775W) profile rising quickly, reaching the ``total'' flux at a radius smaller than the \lya\ (F658N) CoG, which is rising more slowly.  Examining the uncertainties on the CoGs, the difference between the \lya\ and rest-frame UV is at a $>$ 1 sigma significance for much of the profile for CH8-2, and $\sim$ 1 $\sigma$ for CH8-1.  The uncertainties on r$_{h}$ are also quite small, with the half-light radii for \lya\ being 2 -- 4 $\sigma$ larger than that for the rest-frame UV for CH8-1 and CH8-2.

Investigating \fig{fig:profile}, it is apparent that the low significance of the CHa-2 detection is hindering our measurement of its CoG, and thus its half-light radius measurement.  Additionally, for the remaining two objects, while their CoGs indicate larger half-light radii in \lya\ than in the rest-frame UV continuum, one will notice that their F658N CoGs continue to increase out to the maximum radius.  This effect is due to the CTE contribution to the background, which is a primarily positive signal caused by the overlapping CTE tails from the plentiful cosmic rays.  It is thus possible that this CTE effect is artificially increasing the radii we measure in the F658N data.

It is thus prudent to examine these data to ensure that the result of larger sizes in the F658N data is a physical effect, and not an artifact of the data.  We have performed a check on our results by measuring the sizes of galaxies that have F658N magnitudes similar to those in our sample, of 24.6 $\leq$ m$_{F658N}$ $\leq$ 25.2, yet have no \lya\ emission.  For this test sample, we also required that the objects be detected at 5 $\sigma$ significance in both F658N and F775W, that 22 $<$ m$_{F775W}$ $<$ 29, and that the difference between the F658N and F606W magnitudes be $<$ 0.1 mag.  Out of our whole F658N-selected catalog, this yielded 67 objects.  We further culled the sample by excluding objects near image edges, as well as highly extended or clumpy objects, leaving a final sample of 28 objects.  We measured the sizes of these objects in a similar manner as the LAEs in our main sample.  We found the median of the ratio of r$_{h,F658N}$/r$_{h,F775W}$ to be 1.15, with a standard deviation of 0.33.  However, the uncertainty on the radii is much higher in the higher ratio objects; thus we computed a weighted mean, finding r$_{h,F658N}$/r$_{h,F775W}$ = 1.08 $\pm$ 0.02.  This analysis shows that there is a slight systematic effect increasing the radii for objects in the F658N images over the F775W images.  However, with the exception of CHa-2 (which has the least significant detection, and thus is the most difficult to make conclusions about), this $\sim$ 10\% effect is small when compared to the ratio of the radii for our three LAEs (1.1, 1.3 and 2.0 for CHa-2, CH8-1 and CH8-2, respectively) and the stack (1.6).  We conclude that larger F658N sizes in our sample are likely real, but a larger sample of LAEs would increase the confidence in our result.

\subsubsection{Stacking Analysis}
\begin{figure}[t]
\epsscale{1.25}
\plotone{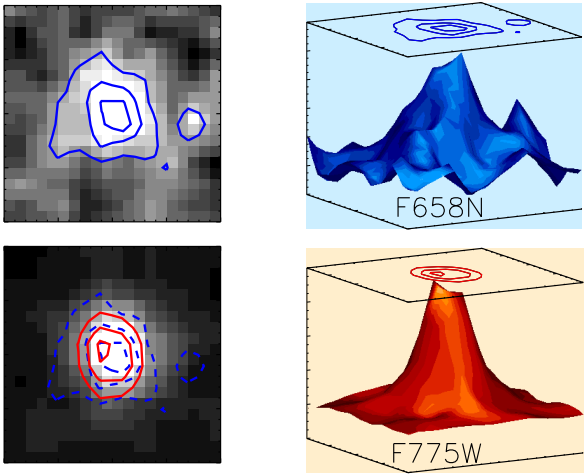}
\caption{Left: Stacked images of the three LAEs in the F658N band (top) and the F775W band (bottom), with both images normalized to their peak flux.  The light in the F658N image is primarily due to \lya\ emission in these galaxies, while the F775W band shows the rest-frame UV continuum emission longward of \lya.  The contours represent regions of constant flux, corresponding to 0.3, 0.5, 0.7 and 0.9 of the peak flux.  The F658N contours are also shown in the F775W image as dashed curves.  Right:  Three-dimensional surface profile of the F658N stack (top) and the F775W stack (bottom).  The contours are the same as in the left-hand panels.  The F775W stack has a steeper profile (and thus a smaller half-light radius), as shown by the more compact surface profile, and the denser contours.}\label{fig:stack}
\end{figure}

\begin{figure}[!t]
\epsscale{1.2}
\plotone{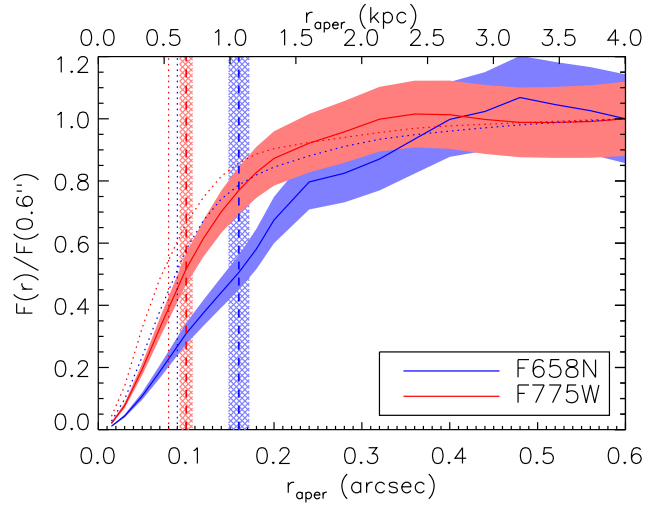}
\caption{Measured light profiles from aperture photometry for the stack of our sample of three spectroscopically confirmed LAEs in blue and red for F658N and F775W, respectively.  The lines, shaded and cross-hatched regions are the same as in \fig{fig:profile}.  Both stacks exhibit the same result as the individual objects -- \lya\ emission appears more extended than the rest-frame UV continuum.}\label{fig:flux_stack}
\end{figure}

As is shown in \fig{fig:stack}, stacking the objects helps to reduce the CTE-affected background.  We see the same results in the stacking analysis in \fig{fig:flux_stack} as hinted at in the individual objects, with the CoGs of the rest-frame UV exhibiting a significantly steeper profile than that of \lya, highlighted here by the smaller uncertainties on the profile due to the greater signal-to-noise of the stacked images.  Similar to the individual results, the \lya\ half-light radius of the stack of LAEs is significantly greater than that of the rest-frame UV continuum, at $\sim$ 5 $\sigma$ significance.  Also of note is that both CoGs reach a value of 1 by $\sim$ 0.4\arcs, and oscillate around 1 (due to image noise) at higher radii.  This implies that our derived half-light radii do not depend on our choice of a maximum radius.  We verified this, as changing the maximum radius from 0.4 -- 0.8\arcs\ changed the resultant half-light radii by less than 1 $\sigma$.

 \fig{fig:stack} highlights this result, showing the two stacked images with contours of constant brightness at 30, 50, 70 and 90\% of the peak flux.  In the right-hand panels, we show three-dimensional surface brightness profiles of these images.  As is evident to the eye, the contours on the \lya\ image are more loosely packed, and the \lya\ 3D image exhibits a broader slope than that of the rest-frame UV continuum image.  Though the difference is slight, primarily due to the faint nature of these objects and the difficulty of space-based narrowband observations, these results are significant.

\subsubsection{Potential Future Improvements}
After the completion of our analysis, we were made aware of potential future improvements to the correction of the poor CTE in the ACS data \citep{anderson10}.  While our analysis shows that the charge trailing is not significantly affecting our size measurements in the F658N data (\S 5.2.2), we were able to reprocess sections 13 and 22 of our F658N data using the updated CTE correction, in order to verify our results (see \fig{fig:thesislaes}).  Briefly, the correction is based on a study of the trails behind warm pixels in dark exposures.  The  algorithm performs a mild deconvolution to restore the flux from the trails into the delta-function warm pixels.  The correction has been demonstrated to work well on  backgrounds greater than 5 electrons, but at the time of  development, sufficient data did not exist to calibrate  the correction for backgrounds below this (the F658N images discussed here have backgrounds much less than this).  Nevertheless, the current algorithm has been shown to correct the majority of CTE blurring, even at essentially zero background\footnote[3]{One additional issue of CTE that enters in at low background is the impact of the read-noise, which did not go through the charge-transfer process.  We examined the corrections with and without the readnoise mitigation employed in \citet{anderson10} and found the resulting images to be essentially the same.}.  In an aperture of radius 0.3\arcs, the CTE-corrected data are $\sim$ 0.2 mag deeper than the uncorrected data, which pushes these {\it HST} narrowband data deeper than the existing ground-based data.  

For our three LAEs, we found F658N sizes in the CTE-corrected data of 0.18\arcs\ $\pm$ 0.02\arcs, 0.21\arcs\ $\pm$ 0.02\arcs\ and 0.21\arcs\ $\pm$ 0.02\arcs\ for CHa-2, CH8-1 and CH8-2, respectively.  This implies that our size measurements for CH8-1 and CH8-2 are likely not adversely affected by the CTE problems, and also that CHa-2 may in fact be resolved, and larger in \lya\ than in the rest-frame UV.  Stacking these objects, we find nearly identical results to our uncorrected stack, with r$_{h,F658N}$ = 0.17\arcs\ $\pm$ 0.01\arcs\ and r$_{h,F775W}$ = 0.10\arcs\ $\pm$ 0.01\arcs.

We currently plan to re-process all of our F658N data with the CTE correction, and do a new selection for LAEs based solely on the {\it HST} data to increase our sample of LAEs with resolved \lya\ emission.  However, the CTE-correction algorithm needs to be verified at the low sky levels present in our data, and this requires new dark frames to be obtained.  This work will be presented in a future paper.

\subsubsection{Interpretation of Results}
Our results indidicate that in our sample of LAEs, the \lya\ light is emitted from a larger region than the rest-frame UV continuum light.  This result is intriguing, since both types of photons likely originate in the same location --- the H\,{\sc ii} regions within the galaxy --- thus one may expect both sets of photons to exhibit the same light profiles.  However, \lya\ photons are resonantly scattered by neutral hydrogen, while the rest-frame UV continuum is not.  In an interstellar medium (ISM) that is homogeneous, \textit{if} there is no dust, this resonant scattering will result in a decoupling between the observed location of \lya\ emission and the rest-frame UV continuum emission, with much of the \lya\ emission eventually escaping from a random location far from its origin, appearing as an extended halo.  However, in recent years, we have learned that many LAEs do in fact contain dust \citep[e.g.,][]{pirzkal07, finkelstein08, finkelstein09a, pentericci09}.  In a dusty homogeneous ISM, where dust is evenly mixed with neutral hydrogen, resonant scattering will result in the majority of \lya\ photons being absorbed by dust; thus a pure homogeneous ISM is unlikely, given that these galaxies exhibit \lya\ in emission.

\begin{figure}[!t]
\epsscale{1.1}
\plotone{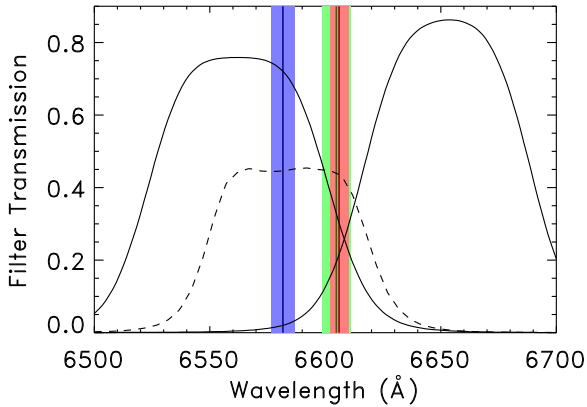}
\caption{Filter profiles of the ground-based NB656 and NB665 filters are shown as the solid black lines, while the {\it HST}/ACS F658N filter profile is shown as the dashed black line.  The position of the \lya\ emission lines for CHa-2, CH8-1 and CH8-2 are shown by the blue, red and green lines, respectively.  The colored shaded regions denote the extent of the emission lines, as measured from the line FWHMs from the observed IMACS spectra.}\label{fig:filt2}
\end{figure}

On the other hand, if the ISM is \textit{inhomogeneous}, \lya\ can still escape in a scattered halo even if dust is present \citep{neufeld91, hansen06, finkelstein07, finkelstein08, finkelstein09a}.  In an idealized case where the ISM is purely clumpy, nearly all \lya\ photons can escape as they scatter off of H\,{\sc i} at the surface of the clumps and are thus screened from the dust.  Even if an ISM is only partially clumpy, this geometry still allows \lya\ to escape, though predominantly in a scattered halo.  Additionally, if the majority of Ly$\alpha$ emission escapes only after scattering off of the receding edge of an outflow, the (now redshifted) Ly$\alpha$ photons would also be decoupled from the rest-frame UV continuum, perhaps appearing in a larger halo as well.

Such halos have been predicted, but have yet to be observed at high-redshift.  Given the modest signal-to-noise of our \lya\ imaging detections, it is likely that we have not detected the full extent of these halos --- rather we are just seeing the tip of the iceberg, in that the \lya\ light appears more extended that the rest-frame UV continuum.  However, given the low signal-to-noise of the LAEs in our data, it is difficult to tell if we are truly seeing the edge of the Ly$\alpha$ emission.  One way to check whether our imaging has captured all of the \lya\ light is to compare the fluxes from the F658N imaging to that from ground-based photometry, which can be more sensitive to diffuse emission given the larger telescope apertures, and reduced sensitivity to read noise.

We can perform this analysis for our sample of LAEs, which have their ground-based narrowband magnitudes tabulated in Table 1 of \citet{finkelstein09a}.  These magnitudes are 24.15 $\pm$ 0.11, 24.44 $\pm$ 0.15 and 24.39 $\pm$ 0.16 for CHa-2, CH8-1 and CH8-2, respectively.  Comparing these magnitudes to those of the same objects from {\it HST} in the F658N data, we find that all objects have significantly greater ground-based narrowband fluxes than from {\it HST}, with flux ratios of f$_{ground}$/f$_{HST}$ of 2.0 $\pm$ 0.6, 2.0 $\pm$ 0.5 and 1.3 $\pm$ 0.3 for these three objects, respectively.  This provides further evidence that these objects have significant \lya\ emission escaping in a diffuse halo, as the ground-based narrowband imaging detects 2$\times$ more flux than {\it HST} for two objects, and 30\% more for one object.  However, this is not quite so straightforward, as the interpretation depends on exactly where the \lya\ flux resides in their respective narrowband filters.  

\fig{fig:filt2} shows the ground based narrowband filters used to select these three objects, as well as that of F658N on ACS.  The colored vertical lines denote the position of \lya\ for these three objects, while the lighter shaded regions denote the full-width at half-maximum of these lines, as measured from the IMACS spectra presented in \S 3.1.  From this figure, we can see that all objects have their \lya\ emission lines encompassed within the FWHM of the F658N filter.  However, only CHa-2 has \lya\ within the FWHM of the NB656 ground-based filter --- both CH8-1 and CH8-2 were detected even though their redshift puts the emission line at $<$ 30\% of the peak F665N filter transmission.  This means that the factor of two flux increase in CHa-2 from the ground over {\it HST} appears real, while the factor of 2 and 1.3 for the other two objects are actually {\it lower limits}, implying that the extended \lya\ halos in these objects may contain a dominant fraction of the total escaping \lya\ flux.

We caution that as these objects are near the image depth limits in both sets of data, there could be zeropoint issues.  As a test, we compared the narrowband fluxes from the ground-based NB665 image (using 2.3\arcs-aperture magnitudes from the catalog from \citet{finkelstein08}) to the F658N data to see if there is a zeropoint offset.  We examined objects in common in both catalogs, computing the mean magnitude difference in bins of 0.5 mag.  From 21 $<$ m$_{F658N}$ $<$ 25, the mean magnitude difference is always $<$ 0.1 mag.  However, there is of course scatter in the individual objects.  At bright magnitudes, this is small, with $\sigma \sim$ 0.3 mag, increasing to $\sigma$ $\sim$ 0.5 mag at m$_{F658N} < 25$.  The magnitude differences for 2/3 LAEs is significant, as their differences are greater than the 1 $\sigma$ uncertainty in the relative zeropoints.  Thus, this flux difference is intriguing, and a similar analysis with a larger LAE sample will provide greater confidence in this effect.

\section{Conclusions}
We have performed high-resolution {\it HST} F658N narrowband imaging over a portion of the GOODS CDF-S in order to directly image resolved \lya\ emission at $z =$ 4.4.  We have detected \lya\ emission from three spectroscopically confirmed LAEs in these data.

Studying the relative positions of these objects in their \lya\ light (from the F658N data) and their rest-frame UV continuum light (from existing F775W data), we find that our data do not support a positional offset between the two types of emission.  We then measured the light profiles and half-light radii from our three LAEs in both filters.  We find that in all three objects the \lya\ light profile rises more slowly, and has a larger half-light radius than the rest-frame UV continuum emission.  We confirmed this result by stacking the galaxy images in both bands, finding that the \lya\ emission has r$_{h}$ = 1.1 kpc, while the rest-frame UV continuum is more compact with r$_{h}$ = 0.7 kpc.  This implies that the \lya\ light is more spread out, presumably due to effects of resonant scattering, possibly in a clumpy ISM.

To investigate this further, we compared the fluxes of our LAEs in the F658N narrowband to ground-based narrowband measurements, which are more sensitive to diffuse emission due to, among other things, larger telescope apertures.  We find that in all three cases where we have measurements from both ground and space, the ground-based narrowband fluxes are significantly greater than the space-based fluxes, by factors of 1.3 -- 2.0.  This shows that the larger physical sizes detected in the F658N data are only the tip of the \lya\ iceberg, and that the majority of the \lya\ emission may lie in a larger, diffuse halo.  It is thus clearly important to include the ISM geometry and kinematics in any study of Ly$\alpha$ emission at high redshift.

While \lya\ emission is one of the most powerful tools we have to discover and study galaxies at high redshift, the complicated radiative transfer undergone by \lya\ photons in their host galaxies muddle the physics that can be inferred.  In order to maximize our understanding of LAEs and \lya\ emission in general, we need to obtain a greater understanding of how \lya\ makes its way from the H\,{\sc ii} regions where it originates to its point of escape from the galaxy.  Studying the \lya\ spatial profiles provides one estimate of the complex radiative transfer by comparing the \lya\ morphologies to those of the rest-frame UV, but more work is needed to obtain strong detections of these diffuse halos, which likely requires the next generation of ground and space-based observatories.

\acknowledgements
We are grateful to the men and women who worked tirelessly for many years to make Wide Field Camera~3 the instrument it is today, and to the STScI Director M.~Mountain for the discretionary time to make this program possible.  Support for {\it HST} program 11359 was provided by NASA through grants GO-11359.0\*.A from the Space Telescope Science Institute, which is operated by the Association of Universities for Research in Astronomy, Inc., under NASA contract NAS 5-26555.  We wish to thank the astronauts of STS-125 to the {\it Hubble Space Telescope} for risking their lives to bring us a new and improved world class observatory.  Their hard work and dedication is greatly appreciated.  SLF and KDF are supported by the Texas A\&M Department of Physics and Astronomy.  The work of SM and JER is supported by the National Science Foundation grant AST-0808165.


\begin{thebibliography}{33}
\expandafter\ifx\csname natexlab\endcsname\relax\def\natexlab#1{#1}\fi

\bibitem[{{Anderson} \& {Bedin}(2010)}]{anderson10}
{Anderson}, J., \& {Bedin}, L.~R. 2010, MNRAS Accepted, astroph/1007.3987

\bibitem[{{Bertin} \& {Arnouts}(1996)}]{bertin96}
{Bertin}, E., \& {Arnouts}, S. 1996, \aaps, 117, 393

\bibitem[{{Bielby} {et~al.}(2010){Bielby}, {Shanks}, {Weilbacher}, {Infante},
  {Crighton}, {Bornancini}, {Bouch{\'e}}, {Garc{\'{\i}}a Lambas}, {Lowenthal},
  {Minniti}, {Padilla}, {Petitjean}, \& {Theuns}}]{bielby10}
{Bielby}, R., {et~al.} 2010, ArXiv e-prints, astroph/1005.3028

\bibitem[{{Bond} {et~al.}(2010){Bond}, {Feldmeier}, {Matkovi{\'c}}, {Gronwall},
  {Ciardullo}, \& {Gawiser}}]{bond10}
{Bond}, N.~A., {Feldmeier}, J.~J., {Matkovi{\'c}}, A., {Gronwall}, C.,
  {Ciardullo}, R., \& {Gawiser}, E. 2010, \apjl, 716, L200

\bibitem[{{Bond} {et~al.}(2009){Bond}, {Gawiser}, {Gronwall}, {Ciardullo},
  {Altmann}, \& {Schawinski}}]{bond09}
{Bond}, N.~A., {Gawiser}, E., {Gronwall}, C., {Ciardullo}, R., {Altmann}, M.,
  \& {Schawinski}, K. 2009, \apj, 705, 639

\bibitem[{{Ferguson} {et~al.}(2004){Ferguson}, {Dickinson}, {Giavalisco},
  {Kretchmer}, {Ravindranath}, {Idzi}, {Taylor}, {Conselice}, {Fall},
  {Gardner}, {Livio}, {Madau}, {Moustakas}, {Papovich}, {Somerville},
  {Spinrad}, \& {Stern}}]{ferguson04}
{Ferguson}, H.~C., {et~al.} 2004, \apjl, 600, L107

\bibitem[{{Finkelstein} {et~al.}(2009){Finkelstein}, {Rhoads}, {Malhotra}, \&
  {Grogin}}]{finkelstein09a}
{Finkelstein}, S.~L., {Rhoads}, J.~E., {Malhotra}, S., \& {Grogin}, N. 2009,
  \apj, 691, 465

\bibitem[{{Finkelstein} {et~al.}(2008){Finkelstein}, {Rhoads}, {Malhotra},
  {Grogin}, \& {Wang}}]{finkelstein08}
{Finkelstein}, S.~L., {Rhoads}, J.~E., {Malhotra}, S., {Grogin}, N., \& {Wang},
  J. 2008, \apj, 678, 655

\bibitem[{{Finkelstein} {et~al.}(2007){Finkelstein}, {Rhoads}, {Malhotra},
  {Pirzkal}, \& {Wang}}]{finkelstein07}
{Finkelstein}, S.~L., {Rhoads}, J.~E., {Malhotra}, S., {Pirzkal}, N., \&
  {Wang}, J. 2007, \apj, 660, 1023

\bibitem[{{Gawiser} {et~al.}(2006){Gawiser}, {van Dokkum}, {Gronwall},
  {Ciardullo}, {Blanc}, {Castander}, {Feldmeier}, {Francke}, {Franx},
  {Haberzettl}, {Herrera}, {Hickey}, {Infante}, {Lira}, {Maza}, {Quadri},
  {Richardson}, {Schawinski}, {Schirmer}, {Taylor}, {Treister}, {Urry}, \&
  {Virani}}]{gawiser06b}
{Gawiser}, E., {et~al.} 2006, \apjl, 642, L13

\bibitem[{{Gronwall} {et~al.}(2010){Gronwall}, {Bond}, {Ciardullo}, {Gawiser},
  {Altmann}, {Blanc}, \& {Feldmeier}}]{gronwall10}
{Gronwall}, C., {Bond}, N.~A., {Ciardullo}, R., {Gawiser}, E., {Altmann}, M.,
  {Blanc}, G.~A., \& {Feldmeier}, J.~J. 2010, ArXiv e-prints, astroph/1005.3006

\bibitem[{{Hansen} \& {Oh}(2006)}]{hansen06}
{Hansen}, M., \& {Oh}, S.~P. 2006, \mnras, 367, 979

\bibitem[{{Hathi} {et~al.}(2008{\natexlab{a}}){Hathi}, {Jansen}, {Windhorst},
  {Cohen}, {Keel}, {Corbin}, \& {Ryan}}]{hathi08b}
{Hathi}, N.~P., {Jansen}, R.~A., {Windhorst}, R.~A., {Cohen}, S.~H., {Keel},
  W.~C., {Corbin}, M.~R., \& {Ryan}, Jr., R.~E. 2008{\natexlab{a}}, \aj, 135,
  156

\bibitem[{{Hathi} {et~al.}(2008{\natexlab{b}}){Hathi}, {Malhotra}, \&
  {Rhoads}}]{hathi08}
{Hathi}, N.~P., {Malhotra}, S., \& {Rhoads}, J.~E. 2008{\natexlab{b}}, \apj,
  673, 686

\bibitem[{{Koekemoer} {et~al.}(2002){Koekemoer}, {Fruchter}, {Hook}, \&
  {Hack}}]{koekemoer02}
{Koekemoer}, A.~M., {Fruchter}, A.~S., {Hook}, R.~N., \& {Hack}, W. 2002, in
  The 2002 HST Calibration Workshop : Hubble after the Installation of the ACS
  and the NICMOS Cooling System, ed. {S.~Arribas, A.~Koekemoer, \&
  B.~Whitmore}, 337

\bibitem[{{Komatsu} {et~al.}(2010){Komatsu}, {Smith}, {Dunkley}, {Bennett},
  {Gold}, {Hinshaw}, {Jarosik}, {Larson}, {Nolta}, {Page}, {Spergel},
  {Halpern}, {Hill}, {Kogut}, {Limon}, {Meyer}, {Odegard}, {Tucker}, {Weiland},
  {Wollack}, \& {Wright}}]{komatsu10}
{Komatsu}, E., {et~al.} 2010, ArXiv e-prints, astroph/1001.4538

\bibitem[{{Kudritzki} {et~al.}(2000){Kudritzki}, {M{\'e}ndez}, {Feldmeier},
  {Ciardullo}, {Jacoby}, {Freeman}, {Arnaboldi}, {Capaccioli}, {Gerhard}, \&
  {Ford}}]{kudritzki00}
{Kudritzki}, R., {et~al.} 2000, \apj, 536, 19

\bibitem[{{Lai} {et~al.}(2007){Lai}, {Huang}, {Fazio}, {Cowie}, {Hu}, \&
  {Kakazu}}]{lai07}
{Lai}, K., {Huang}, J.-S., {Fazio}, G., {Cowie}, L.~L., {Hu}, E.~M., \&
  {Kakazu}, Y. 2007, \apj, 655, 704

\bibitem[{{Malhotra} \& {Rhoads}(2002)}]{malhotra02}
{Malhotra}, S., \& {Rhoads}, J.~E. 2002, \apjl, 565, L71

\bibitem[{{McLinden} {et~al.}(2010){McLinden}, {Finkelstein}, {Rhoads},
  {Malhotra}, {Hibon}, {Richardson}, {Cresci}, {Quirrenbach}, {Pasquali},
  {Bian}, {Fan}, \& {Woodward}}]{mclinden10}
{McLinden}, E.~M., {et~al.} 2010, ArXiv e-prints, astroph/1006.1895

\bibitem[{{Neufeld}(1991)}]{neufeld91}
{Neufeld}, D.~A. 1991, \apjl, 370, L85

\bibitem[{{Nonino} {et~al.}(2009){Nonino}, {Dickinson}, {Rosati}, {Grazian},
  {Reddy}, {Cristiani}, {Giavalisco}, {Kuntschner}, {Vanzella}, {Daddi},
  {Fosbury}, \& {Cesarsky}}]{nonino09}
{Nonino}, M., {et~al.} 2009, \apjs, 183, 244

\bibitem[{{Oesch} {et~al.}(2010){Oesch}, {Bouwens}, {Carollo}, {Illingworth},
  {Trenti}, {Stiavelli}, {Magee}, {Labb{\'e}}, \& {Franx}}]{oesch10b}
{Oesch}, P.~A., {et~al.} 2010, \apjl, 709, L21

\bibitem[{{Ono} {et~al.}(2010){Ono}, {Ouchi}, {Shimasaku}, {Akiyama}, {Dunlop},
  {Farrah}, {Lee}, {McLure}, {Okamura}, \& {Yoshida}}]{ono10}
{Ono}, Y., {et~al.} 2010, \mnras, 402, 1580

\bibitem[{{Partridge} \& {Peebles}(1967)}]{partridge67}
{Partridge}, R.~B., \& {Peebles}, P.~J.~E. 1967, \apj, 148, 377

\bibitem[{{Pentericci} {et~al.}(2009){Pentericci}, {Grazian}, {Fontana},
  {Castellano}, {Giallongo}, {Salimbeni}, \& {Santini}}]{pentericci09}
{Pentericci}, L., {Grazian}, A., {Fontana}, A., {Castellano}, M., {Giallongo},
  E., {Salimbeni}, S., \& {Santini}, P. 2009, \aap, 494, 553

\bibitem[{{Pirzkal} {et~al.}(2007){Pirzkal}, {Malhotra}, {Rhoads}, \&
  {Xu}}]{pirzkal07}
{Pirzkal}, N., {Malhotra}, S., {Rhoads}, J.~E., \& {Xu}, C. 2007, \apj, 667, 49

\bibitem[{{Retzlaff} {et~al.}(2010){Retzlaff}, {Rosati}, {Dickinson},
  {Vandame}, {Rit{\'e}}, {Nonino}, {Cesarsky}, \& {GOODS Team}}]{retzlaff10}
{Retzlaff}, J., {Rosati}, P., {Dickinson}, M., {Vandame}, B., {Rit{\'e}}, C.,
  {Nonino}, M., {Cesarsky}, C., \& {GOODS Team}. 2010, \aap, 511, A50+

\bibitem[{{Rhoads} {et~al.}(2009){Rhoads}, {Malhotra}, {Pirzkal}, {Dickinson},
  {Cohen}, {Grogin}, {Hathi}, {Xu}, {Ferreras}, {Gronwall}, {Koekemoer},
  {K{\"u}mmel}, {Meurer}, {Panagia}, {Pasquali}, {Ryan}, {Straughn}, {Walsh},
  {Windhorst}, \& {Yan}}]{rhoads09}
{Rhoads}, J.~E., {et~al.} 2009, \apj, 697, 942

\bibitem[{{Shapley} {et~al.}(2003){Shapley}, {Steidel}, {Pettini}, \&
  {Adelberger}}]{shapley03}
{Shapley}, A.~E., {Steidel}, C.~C., {Pettini}, M., \& {Adelberger}, K.~L. 2003,
  \apj, 588, 65

\bibitem[{{Waddington} {et~al.}(1999){Waddington}, {Windhorst}, {Cohen},
  {Partridge}, {Spinrad}, \& {Stern}}]{waddington99}
{Waddington}, I., {Windhorst}, R.~A., {Cohen}, S.~H., {Partridge}, R.~B.,
  {Spinrad}, H., \& {Stern}, D. 1999, \apjl, 526, L77

\bibitem[{{Windhorst} {et~al.}(1998){Windhorst}, {Keel}, \&
  {Pascarelle}}]{windhorst98}
{Windhorst}, R.~A., {Keel}, W.~C., \& {Pascarelle}, S.~M. 1998, \apjl, 494,
  L27+

\bibitem[{{Windhorst} {et~al.}(2010){Windhorst}, {Cohen}, {Hathi}, {McCarthy},
  {Ryan}, {Jr.}, {Yan}, {Baldry}, {Driver}, {Frogel}, {Hill}, {Kelvin},
  {Koekemoer}, {Mechtley}, {O'Connell}, {Robotham}, {Rutkowski}, {Seibert},
  {Tuffs}, {Balick}, {Bond}, {Bushouse}, {Calzetti}, {Crockett}, {Disney},
  {Dopita}, {Hall}, {Holtzman}, {Kaviraj}, {Kimble}, {MacKenty}, {Mutchler},
  {Paresce}, {Saha}, {Silk}, {Trauger}, {Walker}, {Whitmore}, \&
  {Young}}]{windhorst10}
{Windhorst}, R.~A., {et~al.} 2010, ApJS Submitted, astroph/1005.2776

\end{thebibliography}
\end{document}